\documentclass[11pt]{article}
\usepackage{graphicx}
\usepackage{color}
\usepackage{amssymb}
\usepackage{amsmath}
\usepackage{algorithmic}
\usepackage{mhequ}
\usepackage{algorithm}
\usepackage{rotating}
\usepackage{threeparttable}
\usepackage{multirow}
\usepackage{url}
\usepackage{setspace}
\usepackage{amssymb,amsmath,amsthm}
\usepackage{vmargin}
\usepackage{bbm}
\usepackage{amsfonts}
\usepackage{bbm}
\usepackage{amssymb}
\usepackage{amsmath}
\usepackage{amsfonts}
\usepackage{color}
\usepackage{graphicx}
\theoremstyle{plain}
\newtheorem{thm}{Theorem}[section]

\theoremstyle{definition}

\theoremstyle{remark}

\newcommand{\mean}[1]{{\rm E}(#1)}
\newcommand{\var}[1]{{\rm Var}(#1)}
\newcommand{\tr}[1]{ {\rm tr}\left(#1 \right )}
\makeatletter 
\@addtoreset{equation}{section}

\makeatother 
\begin{document}
 \begin{center}
 \huge{On Mean-Variance Analysis \footnote{  Work supported by NSERC grants 371653-09 and MITACS grants 5-26761. We thank the referees for valuable advice, suggestions and a thorough reading of the first version.}}
 \end{center}
 \vspace{0cm}
\begin{center}
{Yang Li \\
Rotman School of Management\\
 University of Toronto \\
  105 St. George Street \\
  Toronto, ON, M5S 3E6\\
 Yang.Li10@rotman.utoronto.ca}
   \end{center}
\begin{center} 
{Traian A.~Pirvu \\
 Dept of Mathematics \& Statistics\\
  McMaster University \\
  1280 Main Street West \\
    Hamilton, ON, L8S 4K1\\
 tpirvu@math.mcmaster.ca}
 \end{center}


\begin{abstract}
This paper considers the mean variance portfolio management
problem. We examine portfolios which contain both primary and derivative securities. The challenge
in this context is due to portfolio's nonlinearities. The delta-gamma approximation is employed to overcome it. Thus, the optimization problem is reduced to a well posed quadratic program. The methodology developed in this paper can be also applied to pricing and hedging in incomplete markets.
\end{abstract}


\vspace{0.5cm}


\section{Introduction}
The main objective in portfolio management is the tradeoff between risk and return.
Markovitz, \cite{Markovitz1} and \cite{Markovitz2} studied the problem of maximizing
portfolio expected return for a given level of risk, or equivalently minimize risk for a given amount of expected return. One limitation of Markovitz's model is that it considers portfolios of primary assets only. 

Recent works looked at the optimal management of portfolios containing primary and derivative assets.
 Here we mention \cite{Rock} and \cite{Cole}. In \cite{Rock}, the author introduces a technique for optimizing CVaR (conditional value at risk) of a portfolio. The paper \cite{Cole} notices that the problem of minimizing CVaR for a portfolio of derivative securities is ill-posed. Furthermore, \cite{Cole} shows that this predicament can be overcome by including transaction costs.
 
 There are some papers  which consider portfolio optimization with non-standard asset classes; we recall \cite{Sir}, \cite{Car}, and \cite{Lo}. In a continuous time model \cite{Sir} looks at the problem of maximizing expected exponential utility of terminal wealth, by trading a static position in derivative securities and a dynamic position in stocks. In a one period model \cite{Car} analyses the optimal investment and
equilibrium pricing of primary and derivative instruments. The paper \cite{Lo} shows how to approximate a dynamic position in options by a static one and this is done by minimizing the mean-squared error.

To the best of our knowledge this paper is the first work to consider the mean variance
Markovitz portfolio management problem in one period model with derivative assets.
 For a portfolio containing many assets (primary and derivatives) the estimation of the correlation matrix is a big
challenge. Practitioners solve this difficulty by projecting portfolios on a smaller numbers of factors. If parametric approaches are used (we work in a multivariate normally distributed returns framework), the projection method creates another problem, since the projections are often nonlinear; in order to overcome it the delta gamma approximation is employed. The delta-gamma approximation is well-known and often used in risk management and portfolio hedging. In the industry practice this approximation does well for small time intervals. By performing the delta gamma approximation the portfolio management problem is reduced to a quadratic program. Another challenge may come from covariance matrix of the factors not being positive definite.  
This issue appeared in some
financial optimization problems; e.g., for portfolios of stocks the sample
correlation matrix is just an approximate correlation (because is usually built from inconsistent data sets)
 and hence not positive definite. This problem is addressed by \cite{Reha} and \cite{Hig}. These
works focus on the extraction of a positive semi-definite variance-covariance matrix, obtained through
the solution of a second-order conic mathematical programming problem. It is a way to convexify  an a priori
non convex problem. In \cite{Reha} and \cite{Hig}, the smallest distortion of the original matrix which
satisfies the desired properties (e.g. being a correlation matrix) is obtained by using Frobenious norm.

The results of our paper can be applied to the problem of pricing and hedging in incomplete markets.
For instance we can consider instruments written on nontradable factors (e.g. temperature) and they
can be hedged with tradable instruments which are highly correlated (this procedure is called cross
hedging). Take as an example weather derivatives (e.g. HDD or CDD); energy prices are considered as the traded correlated instrument (in California a high correlation can be observed
between temperature and energy prices). Perfect hedging is not possible in this paradigm. Minimizing the variance of the hedging error can be captured as a special case of mean variance optimization problem for a 
portfolio of primary and derivative instruments. A survey paper on mean-variance hedging and mean-variance portfolio selection is \cite{Sch}.

Another possible application of our results is the hedging of long maturity instruments with short maturities
ones. As it is well known, the market for long maturity instruments is illiquid, thus the issuers use (static) hedging portfolios of the more liquid short maturity instruments. The interested reader can find out more
about this in \cite{Car1}.

The paper is organized as follows: Section 2 presents the model.
 Section 3 introduces the delta gamma approximation. Section 4 presents the reduction to quadratic programs. Section 5 is an application to pricing and hedging in incomplete markets. 

\section{The Model}

Portfolios returns are derived from the return of individual positions. In practice, it is not
good to model the positions individually because of their correlations. If we have $m$
instruments in our portfolio we would need $m$ separate volatilities plus data on $\frac{m(m-1)}{2}$
correlations, so in total $\frac{m(m+1)}{2}$ pieces of information. This is hard to get for large $m.$

The resolution is to map our $m$ instruments onto a smaller number of $n$ risk factors.
The mapping can be nonlinear (e.g. BS (Black Scholes formula) for option). Let us assume that the factors are represented by a stochastic vector process $S=(S_{1}, S_{2}, \cdots, S_{n}),$
which at all times $t\in(0,\infty)$ is assumed to be of the form 
\begin{equation}\label{tri}
S_t =  \mu t + \Sigma W_{t}.
\end{equation}
Here $\mu$ is the vector of returns, $\Sigma$ is the variance-covariance matrix which is assumed
positive definite, and $W_{t}$ is a standard Brownian motion on a canonical probability space $(\Omega, \mathcal{F}_{t}, \mathcal{F} ).$ The value of portfolio at time $t,$ denoted $V(S,t),$ is of the form

\begin{equation}\label{ret0}
V(S,t) = \sum\limits_{k=1}^{m}x_{k} (t) V_k(S,t),
\end{equation}
where $V_k(S,t),\,\, k=1, \cdots m,$ represents the value of the individual instruments (mapped onto the risk factors),  and $x_{k} (t),\,\, k=1, \cdots m$ stands for the number of shares of instrument $k$ held in the portfolio at time $t.$ We choose the portfolio mix $x_{k} (t),\,\, k=1, \cdots m$ such that the portfolio return $\Delta V$ over time interval $[t, t+\Delta t ]$
\begin{equation}\label{ret}
\Delta V = V(S+\Delta s,t+\Delta t)-V(S,t),
\end{equation}
is optimized in a way which is described below. It turns out to be more convenient to work with the vector of actual proportions of wealth invested in the different assets. 
Thus, at time $t\in (0, \infty),$ we introduce the portfolio weights $w_{k} (t), \,\, k=1, \cdots m,$ by
  
\begin{equation}\label{ret} 
 w_{k} (t)= \frac{x_{k} (t)}{V(S,t)}, \,\, k=1, \cdots m.
 \end{equation}
 In the following we posit the Markowitz mean-variance type problem; given some exogenous
 benchmark return $ r_e(t),$ at time $t$ an investor wants to choose among all portfolios having the same return $ r_e(t),$
 the one that has the minimal variance $\var{\Delta V}:$
\begin{eqnarray*}\label{NMV}
\mbox{(P1)}\qquad \min_{w} & &\var{\Delta V} \nonumber \\
{\rm such\,\, that} & &\mean{\Delta V}  =  r_e(t), \\
	      &	 & \sum_{k=1} ^{m}  w_{k}(t)  V_{k} (S,t)= 1. \nonumber
\end{eqnarray*}
Another possible portfolio management problem is to choose the portfolio with the minimal variance:
\begin{eqnarray*}\label{NMV}
\mbox{(P2)}\qquad \min_{w} & &\var{\Delta V} \nonumber \\
& & \sum_{k=1} ^{m}  w_{k}(t) V_{k} (S,t)  = 1. \nonumber
\end{eqnarray*}

\noindent There are some difficulties in solving $\mbox{(P1)}$ and $\mbox{(P2)}.$ First, we might be short of $\Delta V$ moments information.  Because $\Delta V$  nonlinearly depends on the change of factors, it is not obvious what distribution $\Delta V$ would follow even  if we perfectly learn the p.d.f of $\Delta S$. The situation would not get much better if we only require the moment information of $\Delta V$. The integration for moments might be still hard to calculate explicitly. One way out of this predicament is to use delta gamma approximation. 


\section{Delta-Gamma Approximation} 
The delta-gamma approximation states that a portfolio change during a given time period resulting from the change of underlying factors can be approximated by some second order polynomial function, the coefficients of which are given by the portfolio's sensitivities such as the delta, gamma and theta. It is an important tool in risk management and hedging; for instance, to hedge a portfolio of derivatives with respect to the underlying's change, the delta-gamma approximation is employed to match sensitivities of the portfolio with
those of the hedge instruments.  

Mathematically speaking, this approximation is a second order Taylor expansion of the portfolio change $ \Delta V$
over the time interval $[t, t+\Delta t]:$
\begin{equation}\label{DeltaGamma}
 \Delta V\approx \  \delta V= \frac{\partial V}{\partial t}\Delta t +\delta^{T}  \Delta S +\frac{1}{2}  \Delta S^{T}\Gamma \Delta S ,
 \end{equation}
 where
 $$\delta_{i}=\frac{\partial V }{\partial S_{i}},\qquad \Gamma_{ij}=\frac{\partial^{2}V}{\partial S_{i} \partial S_{j}},\quad i=1,\cdots n.$$
 Since  $$V(S,t) = \sum\limits_{k=1}^{m} x_k(t) V_k(S,t),$$ then
 \begin{equation}\label{e1}
\delta_{i}=\frac{\partial V }{\partial S_{i}} = \sum_{k=1}^{m} x_k (t) \delta^{k}_{i},\,\,\delta^{k}_{i}:=\frac{\partial V_k }{\partial S_{i}},\,\,  i=1,\cdots n,\,\, k=1,\cdots m,
\end{equation} 
\begin{equation}\label{e2}
\Gamma_{ij}=\frac{\partial^{2}V}{\partial S_{i}\partial S_{j}}=\sum_{k=1}^{m} x_k(t)\Gamma^{k}_{ij},\,\, \Gamma^{k}_{ij}:=\frac{\partial^{2}V_k}{\partial S_{i}\partial S_{j}},\,i=1,\cdots n,\,  j=1,\cdots n,
 \, k=1,\cdots m.
 \end{equation}
 It is well known that this approximation performs well as long as the time interval $\Delta t$ is not too big. At this point we formulate the approximated versions of $\mbox{(P1)}$ and $\mbox{(P2)},$ as follows:
 \begin{eqnarray*}\label{NMV}
\mbox{(P3)}\qquad \min_{w} & &\var{\delta V} \nonumber \\
{\rm such\,\, that} & &\mean{\delta V}  =  r_e(t), \\
	      &	 & \sum_{k=1} ^{m}  w_{k}(t) V_{k} (S,t)  = 1, \nonumber
\end{eqnarray*}

\begin{eqnarray*}\label{NMV}
\mbox{(P4)}\qquad \min_{w} & &\var{\delta V} \nonumber \\
& & \sum_{k=1} ^{m} w_{k}(t) V_{k} (S,t)  = 1. \nonumber
\end{eqnarray*}
The next step is to reduce  $\mbox{(P3)}$ and $\mbox{(P4)}$ to quadratic programs and this is done in the next section.

 \section{Quadratic Programs} 

Let us first consider the case of one asset, $m=1.$ In the light of \eqref{tri}, $\Delta S \sim  \mathcal{N}(\mu,\Sigma \sqrt{\Delta t})$.
For computational convenience we assume $\mu$ is the zero vector and $\Delta t=1$.
Next, replace the vector of correlated normals, $\Delta S,$ 
with the vector of independent normals $Z \sim \mathcal{N}(0,I).$
This is done by setting
 $$\Delta S=CZ\qquad \mbox{with}\qquad CC^{T}=\Sigma. $$
  In terms of $Z,$ the quadratic approximation of $\Delta V$ becomes
 $$ \Delta V \approx \delta V= a+(C^{T}\delta)^{T}Z+\frac{1}{2} Z^{T} (C^{T} \Gamma C)Z, $$
 with 
 \begin{equation}\label{a}
 a= \frac{\partial V}{\partial t}\Delta t.
 \end{equation}
  At this point it is convenient to choose the matrix $C$ to diagonalize the quadratic term in the above expression and this is done as follows. Let $\tilde{C}$ be a square matrix such that
 \begin{equation}\label{11}
  \tilde{C}\tilde{C}^{T}=\Sigma
  \end{equation}
   (e.g., the one given by the Cholesky factorization).
 The matrix $\frac{1}{2}\tilde{C}^{T} \Gamma \tilde{C}$ is symmetric and thus admits the representation
 \begin{equation}\label{22}
 \frac{1}{2}\tilde{C}^{T} \Gamma \tilde{C}=U \Lambda U^{T},
 \end{equation}
 where $ \Lambda={\rm diag} (\lambda_{1},\cdots, \lambda_{n}),$ and $U$ is an orthogonal matrix such that $U U^{T}=I.$ 
 Next, set $C=\tilde{C}U$ and observe that
 \begin{equation}\label{^}
 C C^{T}=\tilde{C} U U^{T} \tilde{C}^{T}=\Sigma ,
 \end{equation}
 
 $$\frac{1}{2}{C}^{T} \Gamma {C}=\frac{1}{2} U^{T}(\tilde{C}^{T} \Gamma \tilde{C})U=U^{T}(U \Lambda U^{T})U=\Lambda. $$
 Thus, with 
 \begin{equation}\label{b1}
 b=C^{T}\delta,
 \end{equation}
  we get
 $$ \Delta V \approx \delta V= a+b^{T}Z +Z^{T} \Lambda Z:=Y. $$

 \subsection{Moment Generating Function}
 In this subsection, we explore the moment generating function of  $Y$ and further derive the mean and variance of $Y$. In the light of
 \begin{eqnarray}
 Y&=&\sum_{i=1}^{n} (\lambda_{i} Z^{2}_{i}+b_{i} Z_{i})+a\\&=& \sum_{i=1}^{n}\lambda_{i}\left ( Z_{i} + \frac{b_{i}}{2 \lambda_{i}}\right)^{2}+a-\sum_{i=1}^{n}  \frac{b^{2}_{i}}{4 \lambda_{i}},
 \end{eqnarray}
 it follows that the random variable $Y$ is student distributed, being (up to a constant) the sum of squared independent normally distributed random variables. Thus, it is well known that 
 \begin{equation}\label{eta}
 {\rm E}(\theta Y) =  \exp(\eta(\theta)),
 \end{equation}
where
\begin{equation}\label{mgf}
\eta (\theta)= a\theta +\sum_{j=1}^{n} \eta_{j}(\theta)=a\theta +\sum_{j=1}^{n}\frac{1}{2}\left( \frac{\theta^{2}b_{j}^{2}}{1-2\theta\lambda_{j}}-\log{(1-2\theta \lambda_{j})}  \right) ,
\end{equation}
for all $\theta$ satisfying $\max_{j} \theta \lambda_{j}<\frac{1}{2}$. Direct computations lead to     $$
\begin{array}{rcl}
\displaystyle \frac{d\left ( e^{\eta(\theta)} \right )}{d\theta}  & = &
\exp(\eta(\theta)) \displaystyle \frac{d\eta}{d\theta} \\
& & \\
& = & \exp(\eta(\theta)) \left [ a+\displaystyle \frac{1}{2} \sum\limits_{j=1}^n \left ( \displaystyle \frac{2\theta b_j^2(1-2\theta \lambda_j)-\theta^2 (-2\lambda_j) }{(1-2\theta \lambda_j)^2} - \displaystyle \frac{-2\lambda_j}{1-2\theta \lambda_j}
\right )  \right ]
\end{array}
$$
and
$$
\begin{array}{l}
\displaystyle \frac{d^2\left ( e^{\eta(\theta)} \right )}{d\theta^2}\displaystyle
=  \exp(\eta(\theta)) \left [ a+\displaystyle \frac{1}{2} \sum\limits_{j=1}^n \left ( \displaystyle \frac{2\theta b_j^2(1-2\theta \lambda_j)-\theta^2 (-2\lambda_j) }{(1-2\theta \lambda_j)^2} - \displaystyle \frac{-2\lambda_j}{1-2\theta \lambda_j}
\right )  \right ]^2\\
\hspace{2cm}+\exp(\eta(\theta)) \left [ \frac{1}{2} \sum\limits_{j=1}^n \left ( 2\lambda_j \displaystyle \frac{-2\lambda_j}{-(1-2\theta \lambda_j)^2} + \right . \right .\\
\left . \left .
\displaystyle \frac{(2b_j^2-8\theta b_j^2\lambda_j+4\theta b_j^2\lambda_j)(1-2\theta \lambda_j)^2-(2\theta b_j^2(1-2\theta \lambda_j)+\theta^2 b_j^2 2 \lambda_j)((-2\lambda_j)2(1-2\theta \lambda_j))}{(1-2\theta \lambda_j)^4}  \right ) \right ]
\end{array}
$$
Thus, the first and second moments of $Y$ are 
$$
{\rm E}(Y) = \frac{d\left ( e^{\eta(\theta)} \right )}{d\theta}  \bigg|_{ \atop \theta=0}  = a + \sum _{j=1}^{n}\lambda_j
$$
and
$$
{\rm E}(Y^2) = \frac{d^2\left ( e^{\eta(\theta)} \right )}{d\theta^2}\displaystyle \vline_{\atop \theta=0} = (a + \sum _{j=1}^{n}\lambda_j)^2+\sum _{j=1}^{n}(b_j^2+2\lambda_j^2).
$$
Hence,
$$
{\rm Var}(Y) = {\rm E}(Y^2) - {\rm E}^2(Y) = \sum _{j=1}^{n}(b_j^2+2\lambda_j^2).
$$
In order to ease the notations we assume that $V(S,t)=1,$ so the vector of shares $x$ equals the vector of proportions $w$
(also notice that for simplicity we dropped the $t$ dependence of $w$). We would like to express the mean and variance of $Y$ in terms of  $x$. In the light of \eqref{22}, \eqref{11}, \eqref{e2} and trace properties it follows that

\begin{eqnarray} \label{mean}
{\rm E}(Y)  =  a + \sum _{j=1}^{n}\lambda_j
		 &=&  a+ \tr{U\Lambda U^T}\\\notag
		 &=&  a+ \frac{1}{2} \tr{\tilde{C}^T \Gamma \tilde{C}}
		\\\notag &=&   a+\frac{1}{2} \tr{\sum_{j=1}^{{m}} x_j \Gamma_j \Sigma}
		\\\notag &=&  a+x^Tp,
\end{eqnarray}
where the vector $p$ is defined by $$p:=\frac{1}{2}\left (\tr{\Gamma^1 \Sigma}, \tr{\Gamma^2 \Sigma}, \dots, \tr{\Gamma^m \Sigma} \right )^T.$$
As for the variance, recall that with $b$ of \eqref{b1} it follows that (see \eqref{e1} and \eqref{^})
\begin{equation}\label{b}
\sum _{k=1}^{{n}}b_k^2 = b^Tb = (C^T\delta)^TC^T\delta = \delta^TC^TC\delta = \frac{1}{2}x^T \hat{\Sigma} x,
\end{equation}
where 
$$
 \hat{\Sigma}=2M^T\Sigma M {\rm \ and \ }  M=(M_{ij})=(\delta^k_i), \ i=1,\dots, n, \ k=1, \dots,m.
$$
The matrix $\hat{\Sigma}$ is positive semidefinite. In the light of \eqref{^} and trace properties it follows that

$$
\begin{array}{rcl}
\sum\limits_{k=1}^{{n}} \lambda_j^2 & = & \displaystyle \frac{1}{4}\tr{(C^T\Gamma C)^T (C^T\Gamma C)}\\
						  & & \\
						  & = & \displaystyle \frac{1}{4}\tr{\Gamma C C^T\Gamma C C^T}\\
						  & & \\
						  & = & \displaystyle \frac{1}{4}\tr{\Gamma \Sigma \Gamma \Sigma}\\
						  & & \\
						  & = & \displaystyle \frac{1}{4} \tr{\left (\sum_{j=1}^{{m}} x_j \Gamma^j \Sigma \right )^2}						  \\
						  & = & \displaystyle \frac{1}{4} \left ( \sum_{j=1}^{{m}} x_j^2 \tr{(\Gamma^j)^2\Sigma^2} + 2 \sum_{i \neq j} x_ix_j \tr{\Gamma^i\Sigma \Gamma^j \Sigma} \right ) \\
						  & = & \displaystyle \frac{1}{4} x^T Q x,
\end{array}
$$
where the matrix $Q$ is defined by
 \begin{equation}\label{ww}
 Q_{ij} = \tr{\Gamma^i \Sigma \Gamma^j \Sigma},\,\, i=1,\cdots {m},\,\,  j=1,\cdots {m}. 
 \end{equation} Therefore, we end up with
\begin{equation}\label{variance}
\var{Y} =  \sum _{j=k}^{m}(b_k^2+2\lambda_k^2) = \frac{1}{2}x^T( \hat{\Sigma}+Q) x.
\end{equation}
Thus, from \eqref{mean} and \eqref{variance}, the portfolio problem  $\mbox{(P3)}$ (recall that $x=w$)  becomes
\begin{eqnarray*}\label{NMV-Q}
\mbox{({P}5)}\quad \min_{x} & &\frac{1}{2}x^T( \hat{\Sigma}+Q) x \nonumber \\
{\rm s.t.} & &a+x^Tp =  r_e, \\
	      &	 & \sum_{k=1}^{{m}} V_{k} (t,S)  x_{k} = 1, \nonumber
\end{eqnarray*}

and  $\mbox{(P4)}$ becomes

\begin{eqnarray*}\label{NMV-Q}
\mbox{({P}6)}\quad \min_{x} & &\frac{1}{2}x^T( \hat{\Sigma}+Q) x \nonumber \\
{\rm s.t.}  & & \sum_{k=1}^{{ m}} V_{k} (t,S)  x_{k} = 1. \nonumber
\end{eqnarray*}
 It turns out that the problem $\mbox{({P}5)}$  has a  similar form with the classical mean variance portfolio problem, a quadratic objective function and linear constrains. Notice that the matrix $ \hat{\Sigma}+Q$ is positive definite. This comes from $\frac{1}{2}x^T( \hat{\Sigma}+Q) x= \var{\delta V}>0.$
We wrap up our findings in the following Theorem. 
\begin{thm}\label{optimality}
$\mbox{({P}3)}$ is equivalent to $\mbox{({P}5)},$ which is a convex quadratic program, and thus solvable in polynomial time. 
\end{thm}

\section{Quadratic Hedging}
The results we established so far can also be applied to hedging. The motivation comes from incomplete markets. Indeed, financial markets are fundamentally incomplete. It is well known
 that in incomplete markets perfect hedging is not possible. One
  way to solve this problem is to consider quadratic hedging; that
  is, minimize the variance of the hedging error. Let $F$ be a payoff
  of the form $F=V_{1}(S_{t+\Delta t}, t+\Delta t),$ for some map $V_{1}.$
  We would like to hedge this payoff by some instruments which are of the form
  $V_{k}(S, t),\,\, k=2,\cdots,l$ (with $l$ possible less than $n,$ whence the incompleteness).
  For simplicity assume that in this market borrowing and lending of cash is done at zero interest
  rate (this can be easily achieved if one takes the zero coupon bonds as numeraire). Given the number of shares $(x_{1}, x_{2}, \cdots, x_{l})$ in the hedging portfolio, the hedging error is
  $$-\sum_{k=1}^{l+1} x_{k} \Delta V_{k}(S,t),$$
  with $x_{1}=-1,$ and $ \Delta V_{l+1}(S,t)=1.$ Therefore, the problem of minimizing
  the variance of hedging error is of the form $(\mbox{P2}).$ The initial amount needed to finance the hedging portfolio is 
  $$ x_{l+1}+V_{1}(S_{t}, t).$$



\end{document}